\documentclass[12pt, preprint]{aastex}
\usepackage{apjfonts, emulateapj5, natbib}
\input psfig.tex

\newcommand{\oiii}{{[\sc O\,iii]}}
\def\hb{H$\beta$ }

\newcommand{\feii}{Fe\,{\sc ii}}

\slugcomment{To appear in {\it The Astrophysical Journal Letters.}}

\def\farcs{\hbox{$.\mkern-4mu^{\prime\prime}$}}

\shortauthors{JIANG ET AL.}
\begin{document}

\title{Rapid Infrared Variability of Three Radio-loud Narrow-line Seyfert 1
Galaxies: A View from the {\it Wide-field Infrared Survey Explorer}}
\author{{Ning~Jiang\altaffilmark{1,2}}, Hong-Yan~Zhou\altaffilmark{1,2,3},
Luis~C.~Ho\altaffilmark{4},
Weimin~Yuan\altaffilmark{5}, 
Ting-Gui~Wang\altaffilmark{1,2}, Xiao-Bo~Dong\altaffilmark{1,2},
Peng~Jiang\altaffilmark{1,2},
Tuo~Ji\altaffilmark{3} and Qiguo~Tian\altaffilmark{3}}

\altaffiltext{1}{Key laboratory for Research in Galaxies and
Cosmology, University of Science and Technology of China,
Chinese Academy of Science, Hefei, Anhui 230026, China; jnac@mail.ustc.edu.cn}
\altaffiltext{2}{Department of Astronomy, University of Science and
Technology of China, Hefei, Anhui 230026, China~ }
\altaffiltext{3}{Polar Research Institute of China,451 Jinqiao Road, Pudong,
                 Shanghai 200136, China}
\altaffiltext{4}{The Observatories of the
Carnegie Institution for Science, 813 Santa Barbara Street,
Pasadena, CA 91101, USA}
\altaffiltext{5}{National Astronomical Observatories,
Chinese Academy of Sciences, Beijing 100012, China}

\begin{abstract}
Using newly released data from the {\it Wide-field Infrared Survey Explorer},
we report the discovery of rapid infrared variability in three radio-loud 
narrow-line Seyfert 1 galaxies (NLS1s) selected from the 23 sources in the 
sample of Yuan et al. (2008).  J0849+5108 and J0948+0022 clearly show intraday 
variability, while J1505+0326 has a longer measurable time scale within 180 
days. Their variability amplitudes, corrected for measurement errors, are
$\sim 0.1-0.2$ mag.  The detection of intraday variability restricts the 
size of the infrared-emitting region to $\sim 10^{-3}$ pc, significantly 
smaller than the scale of the torus but consistent with the base of a jet.
The three variable sources are exceptionally radio-loud, 
have the highest radio brightness temperature among the whole sample, and all 
show detected $\gamma$-ray emission in \emph{Fermi}/LAT observations.
Their spectral energy distributions resemble those of low-energy-peaked 
blazars, with a synchrotron peak around infrared wavelengths.
This result strongly confirms the view that at least some radio-loud NLS1s
are blazars with a relativistic jet close to our
line of sight. The beamed synchrotron emission from the jet
contributes significantly to and probably dominates the spectra in
the infrared and even optical bands.
\end{abstract}
\keywords{galaxies: individual (SDSS J084957.98+510829.0, SDSS
J094857.32+002225.5, SDSS J150506.48+032630.8) --- galaxies: active
--- galaxies: jets --- infrared: galaxies}

\section{Introduction}
Active galactic nuclei (AGNs; including Seyfert galaxies and
quasars), powered by accretion onto
supermassive black holes, show multiwavelength variability
on time scales from years to less than a day. As one of the first
recognized properties of quasars, variability has served as an
important tool to investigate the emission processes of AGNs 
(e.g., Ulrich et al. 1997; Peterson 2001). In particular, intraday 
variability can provide essential
constraints on the central regions of AGNs on physical scales
smaller than the Solar System, which in general cannot be resolved directly
by current observational capabilities.  Intraday variability has been
detected throughout nearly the entire observable electromagnetic
spectrum in many radio-loud AGNs with flat radio spectra,
which are often collectively called blazars (see review in
Wagner \& Witzel 1995). Radio-loud AGNs plausibly contain a
relativistic jet originating near the central black hole. Blazars are
believed to be those with a relativistic jet oriented  close
to the line of sight; thus, their nonthermal jet emission is highly
Doppler boosted.

Narrow-line Seyfert 1 galaxies (NLS1s), originally defined by
Osterbrock \& Pogge (1985), are a special class of AGNs in which the
broad permitted lines are relatively narrow and the
\oiii\,$\lambda5007$ emission line is weak (see Pogge 2000 for a
review).
Apart from this, NLS1s also show strong \feii\ emission lines in
their optical/ultraviolet spectra, steep soft X-ray spectra, and
rapid X-ray variability (Boller et al. 1996; Leighly 1999; Sulentic
et al. 2000; Zhou et al. 2006). Zhou \& Wang (2002) found that the
fraction of radio-loud objects among NLS1s ($\sim6\%$) is significantly less than
that found in normal type 1 quasars ($\sim13\%$).  Very radio-loud NLS1s,
those with radio-loudness parameter $R>100$, where $R$ is
commonly defined as the ratio of the radio flux density at 6~cm to optical
flux density at 4400~\AA\ (Kellermann et al. 1989), are extremely rare, 
comprising only $2.5\%$ of the NLS1 population (Komossa et al. 2006).
Of particular interest,
some individual objects are found to exhibit blazar-like behavior
(e.g., Zhou et al. 2003, 2005, 2006; Gallo et al. 2006). Yuan et al.
(2008) presented a comprehensive study of 23 genuine radio-loud NLS1s
(hereinafter the Y08 sample).
The radio sources in radio-loud NLS1s are ubiquitously compact, unresolved on scales
of several arcseconds. Some of them show interesting properties, including 
flat radio spectra, compact VLBI cores, very high radio brightness 
temperatures ($T_{\rm B}$), enhanced optical continuum emission, flat X-ray 
spectra, large-amplitude X-ray flux and spectral variability, 
and blazar-like spectral energy distributions (SEDs). 
Based on recent observations taken by
\emph{Fermi Gamma-ray Space Telescope (Fermi)}, some of the radio-loud NLS1s 
indeed display a hard X-ray component and prominent $\gamma$-ray
radiation, which undoubtedly arise from a relativistic jets pointed toward us
(Abdo et al. 2009a, 2009b; Foschini et al. 2011). The blazar
interpretation of radio-loud NLS1s appears to be successful, but additional evidence 
is still needed for the majority of the class. Flat and inverted
radio spectra are also often seen in some normal Seyfert galaxies,
although most have much less power compared to the objects studied
in the Y08 sample (e.g., Ho \& Ulvestad 2001).

As a general feature of blazars, intraday variability is expected in
radio-loud NLS1s if the above interpretation is true. This has been
confirmed by some optical intra-night 

\begin{figure*}
\centerline{\psfig{file=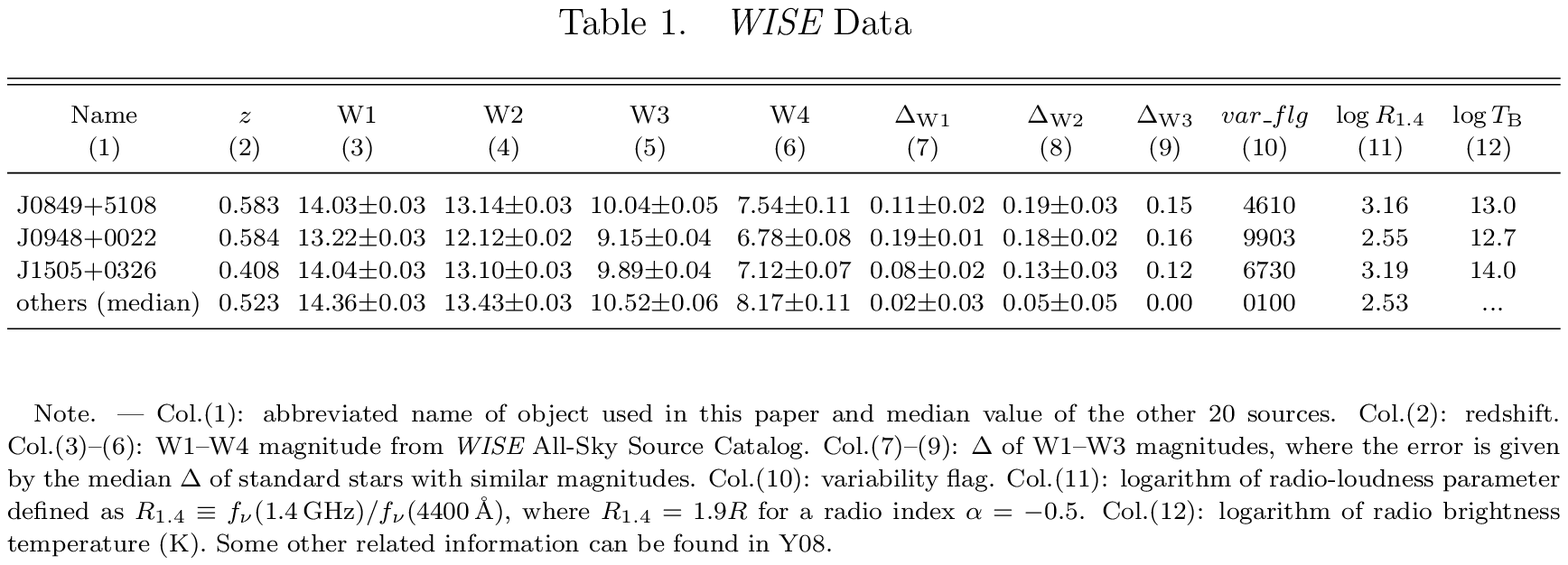,width=18cm,angle=0}}
\end{figure*}
\noindent
observations (e.g., Liu et al.
2009; Shi \& Shan 2011). Extending such an investigation to the infrared (IR) 
becomes widely possible only after the recent mission of \emph{Wide-field 
Infrared Survey Explorer} ({\it WISE}; Wright et al. 2010), which achieves a
sensitivity in the $12\,\mu$m band more than 100 times higher than the
most comparable previous mission \emph{Infrared Astronomical Satellite} 
({\it IRAS}). Indeed, all 23 objects in the Y08 sample are detected by {\it 
WISE} but none of them except one (SDSS J163323.58+471859.0, hereinafter 
J1633+4718)
\footnote{However, the IRAS source is most likely
associated with the star-forming nuclei of a companion galaxy, 
separated by $\sim4\arcsec$ from the AGN (Yuan et al. 2010).}
was detected by {\it IRAS}.  {\it WISE} has mapped the entire sky
in four bands centered at 3.4, 4.6, 12, and 22 $\mu$m
(hereinafter the W1, W2, W3, and W4 bands) with angular resolutions of 
6\farcs1, 6\farcs4, 6\farcs5, and 12\arcsec, respectively. 
The field-of-view of {\it WISE} is
$47^\prime\times47^\prime$ and has a small (10\%) overlap between
adjacent fields in one orbit. The scan circle advances by about
4$^\prime$ per orbit. Thus, there are typically 12 successive
orbits covering a given source. With $\sim 15$ orbits per day,
the observing cadence of {\it WISE} is well suited for studying
intraday variability. It is worth noting that the perspective from
IR variability has some unique advantages. First, light in the IR is
much less affected by dust extinction and/or gas absorption than in
the optical, UV, or X-ray bands. Even compared with radio, IR is
free from contamination from refractive interstellar scintillation,
which is wavelength-dependent in the radio but negligible in the IR.

This Letter reports the discovery of rapid IR variability in three radio-loud 
NLS1s selected from the Y08 sample using recently released data from 
{\it WISE}.  We assume a cosmology with $H_{0} =70$ km~s$^{-1}$~Mpc$^{-1}$, 
$\Omega_{m} = 0.3$, and $\Omega_{\Lambda} = 0.7$.

\section{Data Analysis and Results}
The {\it WISE} All-Sky Data Release Source Catalog contains position and
four-band photometric data for 563,921,584 objects detected on the
coadded atlas images (Cutri et al. 2012). Photometry was performed
by fitting point-spread functions (PSFs) simultaneously to all the
individual exposures covering an object. A variability flag,
$var\_flg$, is assigned to every source in each band, giving the probability 
of flux variation as represented by an integer value from 0 to 9.
A value of 0 indicates insufficient or inadequate data to
determine the variability probability, while values from 1 to 9
indicate increasing likelihood of variability (see Hoffman et al.
2012 for details). According to $var\_flg$ in the {\it WISE} catalog,
there are tree radio-loud NLS1s in the Y08 sample with probable variability:
SDSS~J084957.98+510829.0 (hereinafter J0849+5108), SDSS~J094857.32+002225.5 
(hereinafter J0948+0022), and SDSS~J150506.48+032630.8 (hereinafter 
J1505+0326), whose respective variability flags are ``4610,'' ``9903,'' and 
``6730.'' For the remaining 20 sources, the variability flags
are no larger than 2 in any of the four bands. The data for
these sources are presented in Table~1.

\begin{figure*}[t]
\centerline{\psfig{file=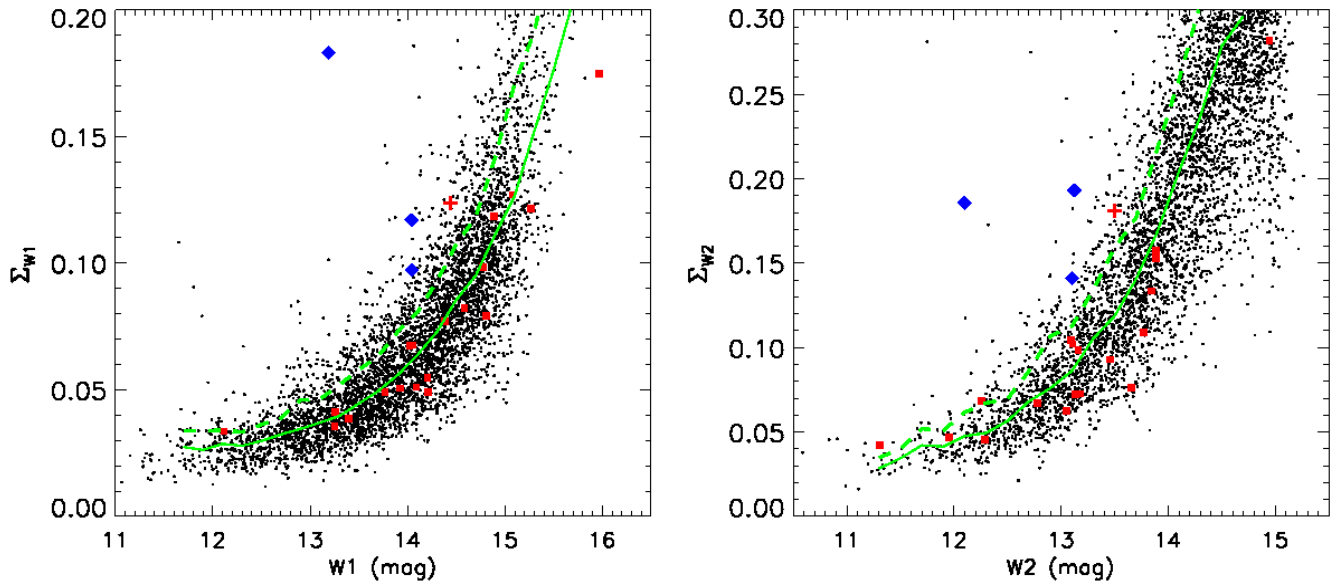,width=18cm}}
\figcaption[rms2m.eps]{
Standard deviation of single-frame photometric measurements as a
function of magnitude for the W1 (left) and W2 (right) band.
Black dots are standard stars from SDSS Stripe 82;
its median values and 1-$\sigma$ upper boundary
are marked with the green solid and dashed lines, respectively.
The three candidate variable sources in the Y08 sample are
plotted with blue diamonds, J1722+5654 as a red plus, and the nonvariable
sources in our sample as red squares.
\label{rms}}
\end{figure*}

We first examine the photometric errors of the {\it WISE}\ data using standard stars
\footnote{http://www.astro.washington.edu/users/ivezic/sdss/catalogs/stripe82.html}
chosen from the SDSS Stripe 82 catalog (Ivezi{\'c} et al. 2007). The
single-frame photometric data can be downloaded from {\it WISE} All-Sky
Single Exposure (L1b) Source Table via the GATOR query service of
the NASA/IPAC Infrared Science Archive
\footnote{http://irsa.ipac.caltech.edu/cgi-bin/Gator/nph-dd}. 
We adopt the PSF profile-fit photometric magnitudes  
\footnote{Brief description of
the definition of profile-fit magnitude as well as other related  parameters 
can be found in Cutri et al. (2012).
}.
The magnitude standard deviation $\Sigma$ for a given star in a band
is calculated as
\begin{equation}\label{eq1}
\Sigma = \sqrt{\frac{1}{n-1} \sum\limits_{i=1}^{N}(m_{i} - \langle m
\rangle)^2},
\end{equation}
where $\langle m \rangle$ is the flux-weighted mean.  Figure~\ref{rms} 
displays $\Sigma$ as function of magnitude in W1 and W2 for about 5000 stars.  
We have also overplotted the 23 radio-loud NLS1s.  The three candidate variable 
sources from the Y08 sample, as well as another one (SDSS J172206.03+565451.6, 
hereinafter J1722+5654), have large variance outside the 1-$\sigma$ scatter of 
the $\Sigma$--magnitude distribution; these outliers are either intrinsically 
variable or their observations have large systematic errors.

We begin by extracting the light curves to reexamine the variability of the 
three candidate variable objects. The signal-to-noise ratio (S/N) is high 
enough to give reliable photometric measurement with $\rm S/N>10$ in W1 and W2 
and $\rm S/N>5$ in W3 for all exposures. Judging by the reduced chi-square 
($\chi^2$) of the profile-fit photometry, which is close to 1, the PSF fitting 
is reliable and there is little contamination from nearby objects. 
Unfortunately, the W4 data have much lower S/N and larger photometric errors;
some are marked as ``null,'' which means that it is either not measurable 
or only an upper limit 

\noindent
can be placed on the flux.  We exclude the W4 data from 
further analysis.  In light of possible photometric zero point offsets in
different fields, for every source we choose as ``standards'' in the field 
30--50 stars with brightness comparable to that of the source. For each 
standard star, we calculate the differential
magnitude between a single observation and the flux-weighted mean
magnitude. Then we take the zero point offset for a given field as the
mean differential magnitude of all the standard stars in the
field. The offsets in W1 and W2 are tiny, typically $<$0.02 mag with a
systematic error of $\sim0.01$ mag.  Most of the standard stars are too
faint in W3, and thus we could not determine a zero point offset for this band

The variability amplitude is commonly measured by the variance of the
observed magnitudes, with the contribution from measurement errors
subtracted. We adopt a formalism similar to that used in Ai et al.
(2010; see also Sesar et al. 2007), whereby
\begin{equation}\label{eq2}
\Delta =\left\{
\begin{array}{ll}
    (\Sigma^2-\xi^2)^{1/2}, & {\rm if~\Sigma > \xi,}\\
    0, & {\rm otherwise} .\\
\end{array}
\right.
\end{equation}
Here the measurement error $\xi$ includes both the 1-$\sigma$ profile-fit 
photometric error and the systematic error from zero point offsets 
($\xi_{\rm zero}$ for W1 and W2, which is, in any case, almost negligible):
\begin{equation}\label{eq3}
\xi^2 = \frac{1}{N} \sum\limits_{i=1}^{N} \xi_{i}^{2} + \xi_{\rm zero}^{2}.
\end{equation}

\begin{figure*}[t]
\centerline{\psfig{file=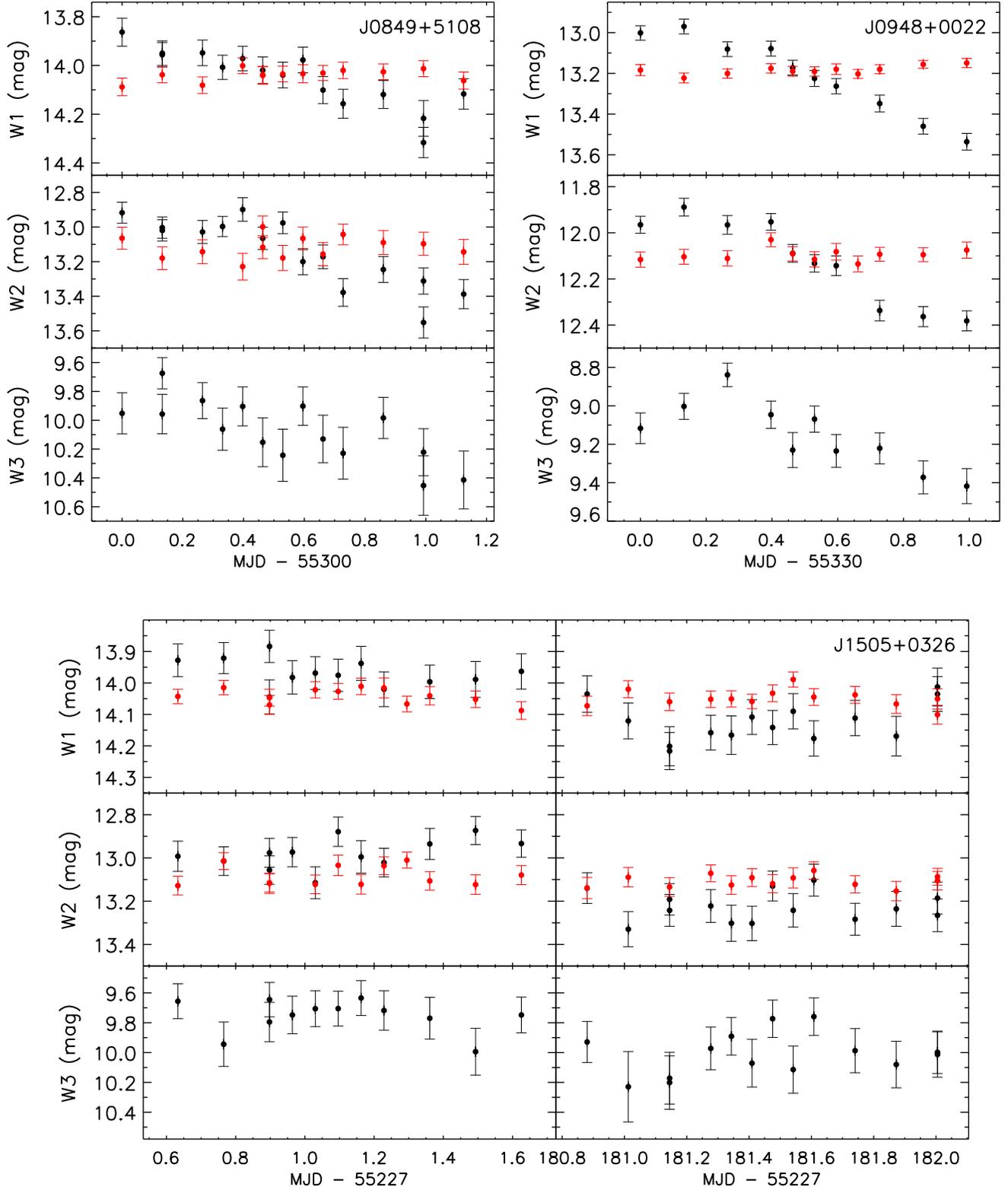,width=18cm}}
\figcaption[lc.eps]{
{\it WISE} light curves constructed from
profile-fit magnitudes; 1-$\sigma$ error bar are
plotted. The red points in the W1 and W2 bands represent
measurements from a typical nearby ``standard'' star, shifted slightly
by a constant for ease of comparison.
\label{lc}}
\end{figure*}

Figure 2 shows the final light curves for the three sources for which 
variability has been detected.  In summary:

\begin{itemize}
{\item {\it J0849+5108} --- There are 15 exposures within 1.2 days, yielding 
$\Delta_{\rm W1}=0.11, \Delta_{\rm W2}=0.19$, and $\Delta_{\rm W3}=0.15$.  As 
a comparison, the standard stars with similar magnitudes in W1 and W2 give a 
median value of $\Delta_{\rm W1}=0.02$ and $\Delta_{\rm W2}=0.03$.}

{\item {\it J0948+0022} --- The 10 exposures within one day give
$\Delta_{\rm W1}=0.19, \Delta_{\rm W2}=0.18$, and $\Delta_{\rm W3}=0.16$.
The standard stars with similar magnitudes yield a median value of 
$\Delta_{\rm W1}=0.01$ and $\Delta_{\rm W2}=0.02$.}

{\item {\it J1505+0326} --- This target was observed during two epochs 
separated by about 180 days. No evident intraday variability can be
seen in either single epoch ($\Delta_{\rm W1,W2,W3} \approx 0$), but the
magnitude jump between the two epochs is considerable (see the
bottom panel of Figure~\ref{lc}). The flux-weighted mean W1 and W2
magnitudes varied by 0.16 and 0.25 between the two epochs, while
the observations for both epochs as a whole yield $\Delta_{\rm
W1}=0.08, \Delta_{\rm W2}=0.13$, and $\Delta_{\rm W3}=0.12$. The W1 and W2
magnitudes are almost equal to those of J0849+5108, and the $\Delta$ of
standard stars is also comparable to that for J0849+5108.  Given the redshift 
of 0.408, the time scale of variability should be shorter than 128
days in the source rest frame.}
\end{itemize}

To verify the accuracy of the variability flag assigned by {\it WISE}, we 
recomputed the variability probability following the formalism devised by 
Hoffman et al. (2012), but using our own error estimates.  With the exception 
of J1633+4718, which shows likely variability according to our test, our 
results are consistent with those of Hoffman et al (2012).  The PSF profile-fit 
magnitudes for J1633+4718, however, are probably unreliable because this 
source is an interacting system of two galaxies separated by $\sim$4\arcsec\ 
as denoted above, which still cannot be resolved by {\it WISE}.  
By contrast, J1722+5654, which 
shows a large magnitude deviation (red plus symbol in Figure~\ref{rms}), has 
$\Delta \approx 0$ after correction for systematic measurement errors.

Lastly, we performed a cross correlation analysis for the three variable 
sources to search for evidence of time lag between the light curves of the 
three bands.  No significant lag was found.

\vskip 0.3cm
\vskip 0.3cm

\section{Discussion and Conclusions}

Using newly released data from {\it WISE}, whose well-designed observation 
mode provides a modest number of repeat observations of the sky, we discovered 
significant IR variability in three out of 23 radio-loud NLS1s selected from 
the sample of Y08.  All three are markedly variable in the W1 (3.4 $\mu$m) 
and/or W2 (4.6 $\mu$m) bands, at greater than $6\sigma$ confidence.  Closer 
examination of their light curves reveals that J0849+5108 and J0948+0022 show 
clear intraday variability, while J1505+0326 

\vskip 0.2cm
\psfig{file=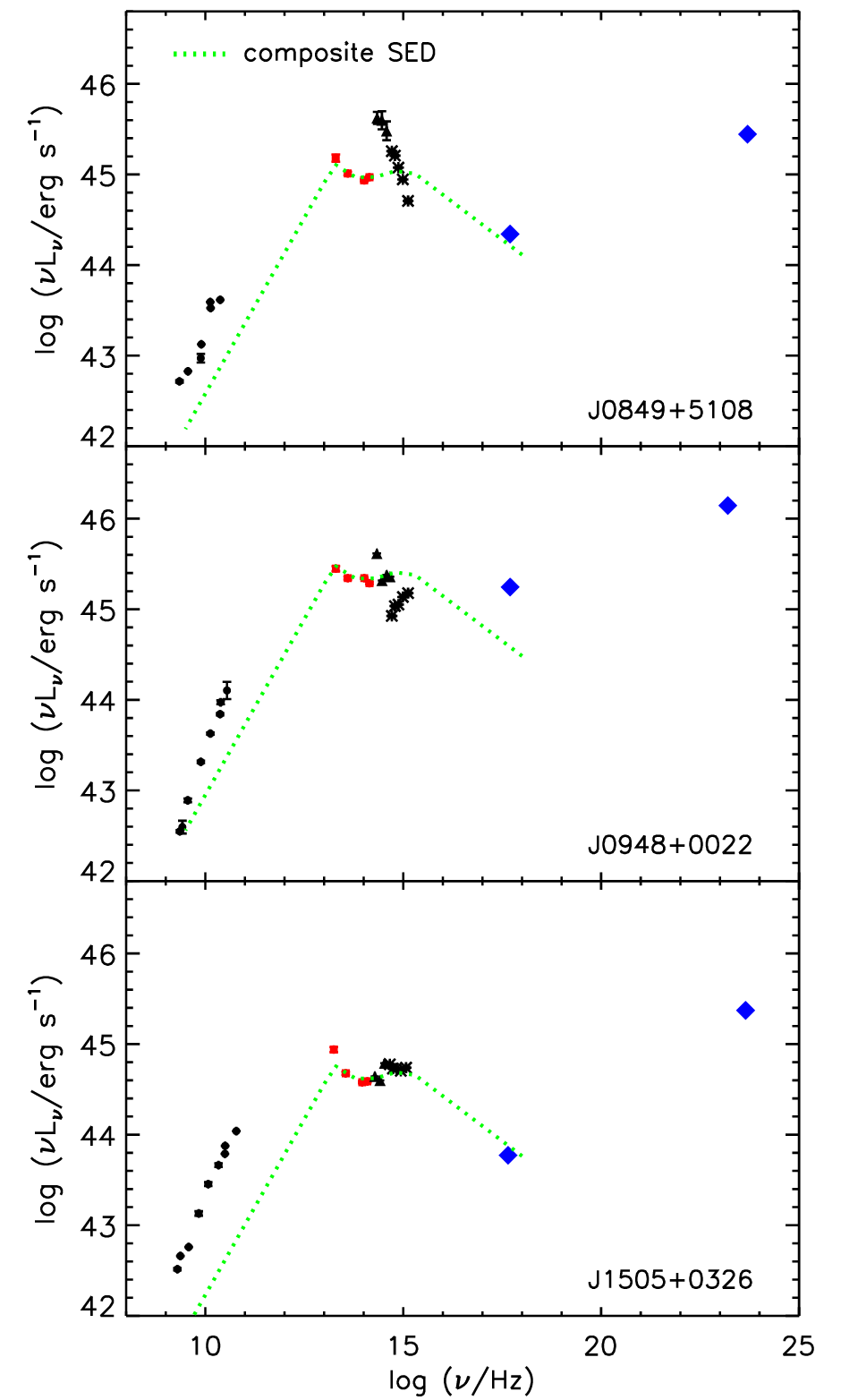,width=8.5cm,angle=0}
\figcaption[sed.eps]{
SEDs of the three variable objects.
Data include radio measurements (black dots), {\it WISE}
photometry (red squares), 2MASS near-IR magnitudes (black triangles),
SDSS $ugriz$ PSF magnitudes corrected for Galactic extinction (black stars),
X-ray and $\gamma$-ray from Abdo et al. 2009b;
D'Ammando et al. 2012 (blue diamonds).
The mean SED of the other 20 nonvariable sources is plotted with a green
dotted line. The radio, 2MASS, and SDSS data were collected from NED.
\label{sed}}
\vskip 0.3cm
\noindent
varies on a longer time scale of 
180 days or less.  Quantifying the variability amplitude using the variance of 
the observed magnitudes (e.g., Ai et al. 2010), after accounting for 
measurement errors, we find $\Delta \approx 0.1-0.2$.  This represents the 
first definitive detection of IR intraday variability for radio-loud NLS1s.

The variability time scale allows us to place limits on the size of the 
IR-emitting region.  In the case of J1505+0326, for which only relatively 
long-term variability can be determined, the size of emission region is 
confined to be $\lesssim0.1$~pc.  On the other hand, the detection of intraday 
variability in J0849+5108 and J0948+0022 implies that their IR emission region 
is very compact, confined to scales of $\sim10^{-3}$~pc, which corresponds to 
only hundreds of Schwarzschild radii assuming $M_{\rm BH} \approx 3 \times 
10^7\,M_\odot$ (Y08).  Thermal IR radiation in AGNs is normally believed to be 
arise mainly from a dusty torus outside of the dust sublimation radius (Laor 
\& Draine 1993), which for graphite grains is about 0.5~pc for these three 
sources.  If the inferred size of the emission region for J0849+5108 and 
J0948+0022 is representative, it implies that the IR radiation comes from a 
spatial scale much smaller than that of the torus, perhaps no larger than that 
of the accretion disk.  This strongly suggests that the IR emission is 
dominated by an additional nonthermal component, presumably associated with 
the jet responsible for the radio emission.  Coincidently, but perhaps not 
surprisingly, in all three objects variable $\gamma$-ray emission has been 
detected recently with \emph{Fermi}/LAT (Abdo et al. 2009a, 2009b; Foschini 
2011; D'Ammando et al. 2012). This confirms the blazar-like nature of these 
three radio-loud NLS1s and supports the hypothesis that they do indeed have 
relativistic jets pointing close to our line of sight.  In this scenario, 
the variable IR emission derives from synchrotron radiation from the base of 
the jet.  

We note, in passing, that intraday variability of J0948+0022 with similar 
amplitude as seen in {\it WISE}\ has been reported also in the optical 
(Liu et al. 2010).  IR observations are, in principle, superior to optical 
observations for the purposes of detecting synchrotron radiation from jets.
IR radiation is less affected by dust extinction and is less contaminated by 
AGN accretion processes that may cause optical variability due to 
instabilities in the accretion flow. 

The fraction of IR-variable sources in the Y08 sample seems to be quite low 
(3/23 or 13\%).  One immediate, trivial reason is probably simply sensitivity. 
About half of the Y08 sample is fainter than the three variable sources, and 
their variability, if present, is likely masked by measurement errors (cf. 
Figure 1).  The degree of radio-loudness also seems to matter.  The majority 
of the parent sample have radio-loudness parameters much lower than those of 
J0849+5108 and J1505+0326.  J0948+0022 has a $R$ parameter 0.6 dex lower than 
the other two variable objects, but it is still higher than half of the full 
sample (see Table~1 and Figure~\ref{sed}).  According to Y08, eight sources 
display prominent radio variability between two epochs separated by several 
years.  The three IR variables discussed here show the largest amplitude of 
radio variations ($\sim40\%-75\%$) and the highest radio brightness 
temperatures, all exceeding the threshold value, usually estimated to be 
$T_{\rm B} \simeq 10^{12}$\,K, above which inverse-Compton catastrophe is 
predicted to occur (Kellermann \& Pauliny-Toth 1969).  Taking the 
inverse-Compton limit of $10^{12}$\,K as a conservative limit, the minimum 
Doppler factors are estimated to be 1.5--4.7, which are within the range of 
values for relativistic jets inferred for classical radio-loud AGNs 
(Ghisellini et al. 1993). 
In addition, they show ubiquitously inverted radio spectra 
and happen to be the flatest three of the 11 sources whose radio spectral
slope have been estimated in Y08, which is consistent with the blazar nature.

Selection by IR variability also must favor objects that are intrinsically 
more IR luminous.  The overall SED of blazars is characterized by two broad 
bumps due to synchrotron emission and inverse-Compton scattering, 
respectively. ``Blazars'' is a collective term for BL Lac objects and 
flat-spectrum radio quasars (FSRQs), which are distinguished from each other 
according to the strength of their emission lines in optical spectra: BL Lac 
objects have very weak emission lines while FSRQs look like normal quasars.  
Depending on the peak frequency of the synchrotron bump, BL Lac objects can be 
divided into low-energy-peaked sources, whose SEDs reach a local maximum at 
IR--optical wavelengths, and high-energy-peaked ones, which peak near the UV 
or soft X-ray band (Giommi \& Padovani 1994).  Similarly, FSRQs can also be 
divided into their high-frequency-peaked and low-frequency-peaked varieties, 
parallel to the classification devised for BL Lac objects (Perlman et al. 
1998).  Our three IR-variable radio-loud NLS1s have SEDs that, in fact, peak 
near the IR (Figure~\ref{sed}) and qualitatively resemble low-frequency-peaked 
FSRQs in terms of their overall energy distribution.  (We note that the 
apparent discontinuity between the IR and optical data points on the SED may 
be caused by variability, since the IR and optical data were not taken 
contemporaneously.) 
This strongly suggests that our IR selection preferentially favors systems with 
IR-peaked SEDs.  

Two tests support the above hypothesis.  First, we examined the IR variability 
properties of the large sample of BL Lac objects from Nieppola et al. (2006).
Judging from the $var\_flg$ parameter from {\it WISE}, 24 out of the 98 
low-energy-peaked sources can be considered variable ($var\_flg>5$) in at 
least one {\it WISE} band, whereas, strikingly, only 6 of 110
high-energy-peaked source qualify as variable.  
As in our sample, it appears that the 
higher incidence of IR variability is related to the higher IR/optical flux 
contribution from jets.  Further evidence comes from examining the strengths
of the optical emission lines.  The median equivalent width of \hb\ for the 
three IR-variable radio-loud NLS1s (23 \AA) is significantly smaller than not 
only the median value of all quasars ($\sim70$~\AA; Zhou et al. 2006; Shen 
et al. 2011) but also that of the parent sample of radio-loud NLS1s (42~\AA).  
This is consistent with the notion that IR-variable sources have a stronger 
optical continuum due to enhanced contribution from a jet.

This study demonstrates that the {\it WISE} All-Sky Survey affords an 
excellent opportunity to study IR variability in AGNs, a powerful diagnostic 
of their radiation mechanism.  Short-term IR variability, in particular, 
is sensitive to spatial scales relevant to the base of the jet, providing 
an effective and efficient probe that complements radio and high-energy 
observations.  We will take full use of the {\it WISE} data to study 
other classes of AGNs in the near future.

\acknowledgements
This work is supported by Chinese Natural Science Foundation through projects 
NSF-10973012, NSF-11033007, SOC project CHINARE2012-02-03 and
Fundamental Research Funds for the Central Universities with grant WK 2030220006.
The research of LCH is supported by the Carnegie Institution for Science.
This publication makes use of data products from the Wide-field Infrared Survey 
Explorer, which is a joint project of the University of California, Los Angeles, 
and the Jet Propulsion Laboratory/California Institute of Technology, 
funded by the National Aeronautics and Space Administration.
This research has made use of the NASA/IPAC Extragalactic Database (NED) which
is operated by the Jet Propulsion Laboratory, California Institute of Technology,
under contract with the National Aeronautics and Space Administration.


\begin{thebibliography}{}
\bibitem[Abdo et al.(2009a)]{2009ApJ...699..976A} Abdo, A.~A., Ackermann, M.,
Ajello, M., et al.\ 2009a, \apj, 699, 976
\bibitem[Abdo et al.(2009b)]{2009ApJ...707L.142A} Abdo, A.~A., Ackermann,
M., Ajello, M., et al.\ 2009b, \apjl, 707, L142
\bibitem[Ai et al.(2010)]{2010ApJ...716L..31A} Ai, Y.~L., Yuan, W., Zhou,
H.~Y., et al.\ 2010, \apjl, 716, L31
\bibitem[Boller et al.(1996)]{1996A&A...305...53B} Boller, Th., Brandt,
W.~N., \& Fink, H.\ 1996, \aap, 305, 53
\bibitem[Cutri \& et al.(2012)]{2012yCat.2311....0C} Cutri, R.~M., Wright, E. L., 
COnrow, T., et al.\ 2012, VizieR Online Data Catalog, 2311
\bibitem[D'Ammando et al.(2012)]{2012arXiv1207.3092D} D'Ammando, F.,
Orienti, M., Finke, J., et al.\ 2012, \mnras, 426, 317
\bibitem[Foschini(2011)]{2011nlsg.confE..24F} Foschini, L.\ 2011,
Narrow-Line Seyfert 1 Galaxies and their Place in the Universe
\bibitem[Gallo et al.(2006)]{2006MNRAS.370..245G} Gallo, L.~C., Edwards,
P.~G., Ferrero, E., et al.\ 2006, \mnras, 370, 245
\bibitem[Ghisellini et al.(1993)]{1993ApJ...407...65G} Ghisellini, G.,
Padovani, P., Celotti, A., \& Maraschi, L.\ 1993, \apj, 407, 65
\bibitem[Giommi \& Padovani(1994)]{1994MNRAS.268L..51G} Giommi, P., 
\& Padovani, P.\ 1994, \mnras, 268, L51 
\bibitem[Ho \& Ulvestad(2001)]{2001ApJS..133...77H} Ho, L.~C.,
\& Ulvestad, J.~S.\ 2001, \apjs, 133, 77
\bibitem[Hoffman et al.(2012)]{2012AJ....143..118H} Hoffman, D.~I., Cutri,
R.~M., Masci, F.~J., et al.\ 2012, \aj, 143, 118
\bibitem[Ivezi{\'c} et al.(2007)]{2007AJ....134..973I} Ivezi{\'c}, {\v Z}.,
Smith, J.~A., Miknaitis, G., et al.\ 2007, \aj, 134, 973
\bibitem[Kellermann \& Pauliny-Toth(1969)]{1969ApJ...155L..71K} Kellermann, K.~I.,
\& Pauliny-Toth, I.~I.~K.\ 1969, \apjl, 155, L71
\bibitem[Kellermann et al.(1989)]{1989AJ.....98.1195K} Kellermann, K.~I.,
Sramek, R., Schmidt, M., Shaffer, D.~B., \& Green, R.\ 1989, \aj, 98, 1195
\bibitem[Komossa et al.(2006)]{2006AJ....132..531K} Komossa, S., Voges, W.,
Xu, D., et al.\ 2006, \aj, 132, 531
\bibitem[Laor \& Draine(1993)]{1993ApJ...402..441L} Laor, A., \& Draine,
B.~T.\ 1993, \apj, 402, 441
\bibitem[Leighly(1999)]{1999ApJS..125..317L} Leighly, K.~M.\ 1999, \apjs,
125, 317
\bibitem[Liu et al.(2010)]{2010ApJ...715L.113L} Liu, H., Wang, J., Mao, Y.,
\& Wei, J.\ 2010, \apjl, 715, L113
\bibitem[Nieppola et al.(2006)]{2006A&A...445..441N} Nieppola, E., Tornikoski,
M., \& Valtaoja, E.\ 2006, \aap, 445, 441
\bibitem[Osterbrock \& Pogge(1985)]{1985ApJ...297..166O} Osterbrock, D.~E.,
\& Pogge, R.~W.\ 1985, \apj, 297, 166
\bibitem[Perlman et al.(1998)]{1998AJ....115.1253P} Perlman, E.~S.,
Padovani, P., Giommi, P., et al.\ 1998, \aj, 115, 1253
\bibitem[Peterson(2001)]{2001sac..conf....3P} Peterson, B.~M.\ 2001,
Advanced Lectures on the Starburst-AGN, 3
\bibitem[Pogge(2000)]{2000NewAR..44..381P} Pogge, R.~W.\ 2000, \nar, 44, 381
\bibitem[Sesar et al.(2007)]{2007AJ....134.2236S} Sesar, B., Ivezi{\'c}, 
{\v Z}., Lupton, R.~H., et al.\ 2007, \aj, 134, 2236 
\bibitem[Shen et al.(2011)]{2011ApJS..194...45S} Shen, Y., Richards, G.~T.,
Strauss, M.~A., et al.\ 2011, \apjs, 194, 45
\bibitem[Shi \& Shan(2011)]{2011PABei..29..452S} Shi, G., \& Shan, H.-G.\ 2011,
Progress in Astronomy, 29, 452
\bibitem[Sulentic et al.(2000)]{2000ApJ...536L...5S} Sulentic, J.~W.,
Zwitter, T., Marziani, P., \& Dultzin-Hacyan, D.\ 2000, \apjl, 536, L5
\bibitem[Ulrich et al.(1997)]{1997ARA&A..35..445U} Ulrich, M.-H., Maraschi, L.,
\& Urry, C.~M.\ 1997, \araa, 35, 445
\bibitem[Wagner \& Witzel(1995)]{1995ARA&A..33..163W} Wagner, S.~J.,
\& Witzel, A.\ 1995, \araa, 33, 163
\bibitem[Wright et al.(2010)]{2010AJ....140.1868W} Wright, E.~L.,
Eisenhardt, P.~R.~M., Mainzer, A.~K., et al.\ 2010, \aj, 140, 1868
\bibitem[Yuan et al.(2008)]{2008ApJ...685..801Y} Yuan, W., Zhou, H.~Y.,
Komossa, S., et al.\ 2008, \apj, 685, 801 (Y08)
\bibitem[Yuan et al.(2010)]{2010ApJ...723..508Y} Yuan, W., Liu, B.~F., 
Zhou, H., \& Wang, T.~G.\ 2010, \apj, 723, 508 
\bibitem[Zhou \& Wang(2002)]{2002ChJAA...2..501Z} Zhou, H.-Y.,
\& Wang, T.-G.\ 2002, \cjaa, 2, 501
\bibitem[Zhou et al.(2003)]{2003ApJ...584..147Z} Zhou, H.-Y., Wang, T.-G.,
Dong, X.-B., Zhou, Y.-Y., \& Li, C.\ 2003, \apj, 584, 147
\bibitem[Zhou et al.(2005)]{2005ChJAA...5...41Z} Zhou, H.-Y., Wang, T.-G.,
Dong, X.-B., Li, C., \& Zhang, X.-G.\ 2005, \cjaa, 5, 41
\bibitem[Zhou et al.(2006)]{2006ApJS..166..128Z} Zhou, H., Wang, T., Yuan,
W., et al.\ 2006, \apjs, 166, 128
\bibitem[Zhou et al.(2007)]{2007ApJ...658L..13Z} Zhou, H., Wang, T., Yuan,
W., et al.\ 2007, \apjl, 658, L13

\end{thebibliography}
\end{document}